# Antropología de la Informática Social: Teoría de la Convergencia Tecno-Social

**Rommel Salas, MBA**
**Universidad del Turabo, Estudios Doctorales**
**Puerto Rico 2016**

**Resumen**

El humanismo tradicional del siglo XX, inspirado por la cultura del libro, se distanció sistemáticamente de la nueva sociedad de la información digital; el Internet y las herramientas de procesamiento de información revolucionaron el mundo, la sociedad en el transcurso de este periodo desarrolló ciertas características adaptativas, basadas en la convivencia (Humano – Maquina), esta transformación establece su base en el impacto de tres segmentos tecnológicos: Los dispositivos, las aplicaciones y la infraestructura de comunicación social, las cuales están envueltas en diversos cambios físicos, conductuales y cognitivos del ser humano; así como el surgimiento de nuevos modelos de influencia y control social, mediante la nueva comunicación ubicua; no obstante en este nuevo proceso de convivencia se desarrollan nuevos modelos como el "pensamiento colaborativo" y el "InfoSharing"; que gestionan la información social bajo tres dimensiones ontológicas *Humano* (h) – *Información* (i) – *Maquina* (m), el cual es la base de un nuevo ecosistema físico-cibernético, donde coexisten y se desarrollan nuevas unidades sociales llamadas "Comunidades virtuales". Esta nueva infraestructura de comunicación y gestión social de información a dado descubierto áreas de vulnerabilidad denominada "Perspectiva social del riesgo", impactando a todas las unidades sociales por medio del vector masivo de impacto (i); El entorno virtual "H + i + M"; y sus componentes, así como el ciclo de vida de la gestión de información social nos permite entender la trayectoria de la integración "Tecno – Social", y estableciendo de la nueva cibernética, dentro de la convergencia de la tecnología con la sociedad y su nuevos retos de convivencia, encaminados en una nueva visión holística y no pragmática, ya que el componente humano (h) en este entorno virtual es el precursor del futuro y necesita ser estudiado no como una aplicación, sino como el eje de una nueva sociedad.

Palabras clave: *Ciberantropología, cibercultura, comunidades virtuales, cibernética*

Abstract

The traditional humanism of the twentieth century, inspired by the culture of the book, systematically distanced itself from the new society of digital information; the Internet and tools of information processing revolutionized the world, society during this period developed certain adaptive characteristics based on coexistence (Human - Machine), this transformation sets based on the impact of three technology segments: devices, applications and infrastructure of social communication, which are involved in various physical, behavioural and cognitive changes of the human being; and the emergence of new models of influence and social control through the new ubiquitous communication; however in this new process of conviviality new models like the "collaborative thinking" and "InfoSharing" develop; managing social information under three Human ontological dimensions (h) - Information (i) - Machine (m), which is the basis of a new physical-cyber ecosystem, where they coexist and develop new social units called "virtual communities ". This new communication infrastructure and social management of information given discovered areas of vulnerability "social perspective of risk", impacting all social units through massive impact vector (i); The virtual environment "H + i + M"; and its components, as well as the life cycle management of social information allows us to understand the path of integration "Techno - Social" and setting a new contribution to cybernetics, within the convergence of technology with society and the new challenges of coexistence, aimed at a new holistic and not pragmatic vision, as the human component (h) in the virtual environment is the precursor of the future and needs to be studied not as an application, but as the hub of a new society.

1. Introducción

> *"La escritura era en su origen la voz de una persona ausente..."; "La cámara fotográfica ha creado un instrumento que retiene las impresiones fugaces visuales, al igual que un disco gramófono que retiene auditivas igualmente fugaces…"* Sigmund Freud: Civilisation and its Discontents 1929.

El artículo científico "Como podríamos pensar" o "*As we may Think*" (1945) de Vannevar Bush como lo menciona (Levy, 2006), es el inicio del Hipertexto y por consiguiente del Word Wide Web, conocido en conjunto a otros servicios como Internet; ninguna persona podía imaginar que cincuenta años después de esta aportación científica, la convergencia entre la sociedad y la tecnología impactaría los patrones y modelos de vida; cambiando su manera de comunicarse (Fei-Yue Wang, 2007).

Los actuales modelos de comunicación, establecen dentro de la vida cotidiana de nuestra sociedad, nuevos patrones de conducta que son paulatinamente modificados bajo procesos de comunicación interpersonal por medio del uso de tecnología, la flexibilidad y ubicuidad de la comunicación digital (Tolosana, 2007), desarrollando una nueva dimensión científica y tecnológica llamada "*Perspectiva Social del Riesgo*", la cual ha tomado auge bajo los nuevos paradigmas tecnológicos de la sociedad actual, ya que con la convergencia "Tecno-Social" la producción de riesgo se ha incrementado y diversificado en todas las áreas Sociales.

2. Planteamiento del Problema

2.1 El humano y sus capacidades adaptativas

> *¿Dónde está la vida que hemos perdido viviendo?*
> *¿Dónde está la sabiduría que hemos perdido en el conocimiento?*
> *¿Dónde está el conocimiento que hemos perdido en la información?*
>
> *T.S. Eliot, "The Rock", Faber & Faber 1934.*

El estudio de la antropología está basado en tres componentes: Evolutivo, comparativo y multidisciplinario (Rivera, 2009), dentro de sus premisas tenemos a las capacidades adaptativas del ser humano hacia los cambios, ya que los seres humanos heredan la tendencia a la adaptación en nuevos ambientes; sean estos por asimilación o acomodación (Woolfolk, 2006) , lo cual nos permite entender algunos fenómenos causados paulatinamente en la sociedad contemporánea en el periodo digital.

Con la llegada de la tecnología como herramienta de gestión de información, la sociedad se transformó con el impacto de tres componentes tecnológicos: Los dispositivos (*hw - hardware*), Las aplicaciones (*sw - software*), el Internet y sus servicios; lo cual establece la base del pensamiento colaborativo y el "InfoSharing". De acuerdo a los estudios realizados por (Nicotera, 1993), la calidad de una relación interpersonal afecta a diferentes áreas físicas y psicológicas; la comunicación efectiva es una variable determinante en una relación interpersonal saludable de acuerdo al estudio de Nicotera, los cuales dentro de la CTS "Convergencia Tecno-Social" son afectados en todos los componentes de las unidades sociales. Por esta razón la comunicación y sus procesos son estudiados en la actualidad rumbo a la compleja y completa integración tecno-social, lo cual ha llegado a impactar notablemente en la sociedad y en algunas áreas de las unidades sociales como: la seguridad pública, la familia, la industria y la educación; de ahí, el desarrollo de "*La perspectiva social del riesgo*" (Tolosana, 2007), la cual ha establecido patrones de impacto en diferentes componentes de estas unidades.

Como ejemplo tenemos el impacto en la seguridad gubernamental; que a nivel mundial han establecido políticas en torno a la seguridad de la información, este es el caso del Gobierno de los Estados Unidos el cual estableció desde 1996 ciertas regulaciones legales enmarcadas en la seguridad y protección de la información basados en "Los Derechos Constitucionales de sus Ciudadanos" como el: "Electronic Communications Privacy Act", "Childrens Online Privacy Protection Act", "USA Patriot Act" y para el 2013 el "Cybersecurity Framework".

Aunque los beneficios son innumerables dentro de los nuevos procesos electrónicos de gestión de información como la: Agilidad en el servicio público, el acceso a la información de banca personal, los nuevos modelos educativos virtuales, los procesos de comercio electrónico, las herramientas colaborativas en la gestión empresarial y mucho más, lo cual ha permitido agilizar procesos minimizando tiempo y costes. El impacto y desarrollo paulatino e insostenible de fenómenos que han venido desarrollándose desde el inicio, fruto de la falta de criterio ambivalente de los órganos de control social hacia el establecimiento de normas, estándares y modelos que faculten no solamente el desarrollo de la variable (m) -Hardware y Software- sino también el completo proceso de convivencia entre el humano y la máquina (h + m); donde este problema derriba en la profundización de la perspectiva social del riesgo o vulnerabilidad social basada en: influencia, control y replica de conductas dentro de las comunidades virtuales.

3. Justificación

Ya en el siglo XX los nuevos procesos productivos y tecnológicos basados en la globalización y modernidad (Barañano, 2010), establecen una brecha que se distancia de los principios hegemónicos de "Cultura individualista"; ya que esta es absorbida por la "globalización", entendiéndose así como "Un fenómeno inevitable en la historia humana que ha acercado el mundo a través del intercambio de bienes y productos, información, conocimientos y cultura" (ONU, 2016). Además, en esta centuria se observa cambios profundos enmarcados en la transformación de la "cultura del libro" y la nueva sociedad de la información digital como lo menciona (Chavarria, 2013).

Este proceso de globalización por medio de la gestión electrónica de información ha desarrollado dos puntos de divergencia: los que aun quieren mantener la identidad cultural, y el otro que busca homogenización basada en la multiculturalidad como lo menciona (Gómez, 2001), este punto de controversia ha ido subsanándose a través del desarrollo de las comunidades virtuales, las cuales establecen un proceso multicultural y homogéneo, basado en una misma perspectiva, con vastos procesos de convivencia y coexistencia.

Dentro de los cambios vertiginosos paulatinos y segmentados en todas las unidades sociales, diferentes factores son copartícipes del mismo como son la influencia, control y replica de conductas dentro de las comunidades virtuales, los problemas en la seguridad de información y la perspectiva social del riesgo, la falta de criterio ambivalente en los órganos de control social hacia el establecimiento de normas, estándares que hubieran permitido una integración "intergeneracional" fácil y organizada dentro del periodo digital.

Por otra parte el desarrollo vertiginoso de nuevas unidades sociales con características distintas, establecidas bajo requerimientos tecnológicos específicos, y a las cuales se les denomina "Comunidades Virtuales", son parte de una nueva generación de unidades sociales que habitan en el ecosistema físico-cibernético existente, este fenómeno social permite establecer nuevas líneas de investigación científica bajo nuevos modelos que van desarrollándose para afrontar a estas brechas sociales, permitiéndonos establecer nuevos procesos para la gestión de información dentro de la investigación científica; este es el caso de la Netnografia, como nuevo método de recolección y análisis de información, esto como base para el desarrollo de nuevos estudios científicos en pertinencia a la Antropología de la Informática Social, el ecosistema Fisico-Cibernetico, la cibercultura y las comunidades virtuales y la información como vector masivo de impacto.

Entender la responsabilidad que tenemos dentro de la CTS, nos permitirá establecer una estrategia de nuevos modelos morfológicos de convivencia (Humano – Maquina), que impacten las nuevas generaciones y transformen las existentes globalmente, más allá de idiomas, monedas, fronteras y razas.

4. Marco Teórico y Conceptual

4.1. La Netnografia como técnica investigativa en la Antropología de la Informática Social

Dentro de todo proceso de investigación científica; las técnicas y herramientas que se usan son de suma importancia; la investigación de campo como lo menciona (Bartis, 2004), "requiere la observación de primera mano, mediante documentación de lo que el investigador observa y escucha en un sitio particular"; por esta razón la técnica de investigación que usa este estudio es el Netnografíco con un enfoque en el método cualitativo, para entender el mismo es importante establecer procesos jerárquicos sociales que parten de un macro concepto de la antropología social, la cual establece la distinción entre identidad individual e identidad colectiva (Barañano, 2010), el cual manifiesta que los "significados culturales, son sobre todo individuales e interculturales"; esta premisa nos lleva a dilucidar atributos que establecen una de las bases de este estudio.

Esta técnica de investigación científica nos permite realizar diferente tipo de estudios dentro del ecosistema físico-cibernético; con una diversidad de énfasis en diferentes problemáticas que afrontan las unidades sociales, entre las cuales podemos mencionar las enfocadas en conductas aprendidas, estos comportamientos están basados en nuevos valores, costumbres, que están dirigidos y establecidos dentro de las comunidades virtuales emergentes (Bowler, 2010). Esta interacción vinculante (humano – información - maquina) o (h + i + m), nos permite entender factores ontológicos de la sociedad y la cibercultura; aunque el énfasis de muchos investigadores es establecer relaciones, correlaciones, regresiones, predicciones y más métodos cuantitativos entre las características de una comunidad virtual y una tradicional; es importante ver más allá de los antiguos paradigmas del paralelismo tecno-social, estableciendo de esta manera un criterio unificado en cuanto a la nueva forma de vida tecno incluyente, que es la base de estudio en la CTS. El individuo al ser miembro de una unidad social, es objeto de estudio de la "etnografía", la cual busca establecer un proceso de "observación" directa (Murillo, 2016), que permita el levantamiento de información para su posterior análisis.

La teoría ambivalente de la cultura y cibercultura; nos muestra; el enfoque que debemos tener hacia el estudio de los fenómenos conductuales humanos dentro de un nuevo estilo de vida, lo cual implica nuevos procesos de comunicación social basados en la relación: H+i+M (Humano – Información – Maquina), que establece un patrón de comunicación aplicable en todas las unidades sociales; por lo tanto, es pertinente continuar con estudios cuantitativos posteriores de la variable moderadora (i) *información* de este marco conceptual, con el cual podremos medir la influencia, impacto, riesgo y hacer futuros estudios de correlación, regresión; para así entender y afrontar específicos factores que influyen en la brecha que enmarca la nueva generación humana.

4.2. El ecosistema físico-cibernético y la CTS

Los espectros oximorónicos como: la realidad artificial, realidad virtual, mundo virtual, o los Neologismos como: el ciberespacio, netnografia, cibercrimen son *artifacts* inherentes al cambio del periodo tipográfico al digital, los cuales nos permiten establecer y proveer nuevos procesos de comunicación (Ellis, 1995), no obstante, esta base epistemológica busca permitirnos entender la completa interacción que existe entre el humano (*h*) y la maquina (*m*); la cual dio como resultado un cambio completo en la manera de interactuar y un nuevo método de comunicación del ser humano por medio del uso inevitable de dispositivos electrónicos en la comunicación interpersonal.

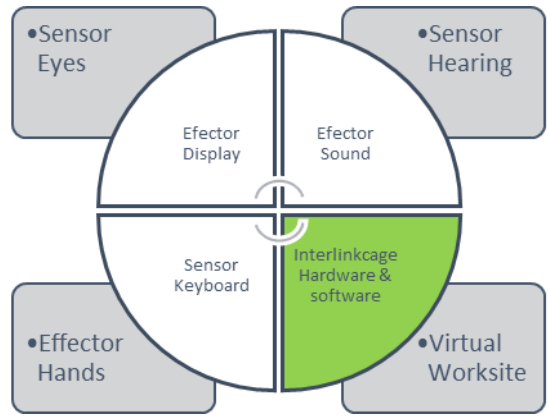

Figura 1 Functional breakdown VE based in Stephen Ellis Model

Este entorno virtual es gestionado por diferentes actores, pero en un mismo ecosistema, y bajo una misma plataforma de comunicación social llamada "Internet"; a lo largo de estos últimos años esta se ha posicionado como la "Infraestructura de comunicaciones y gestión de información", en este estudio no analizaremos información estadística u otros modelos cuánticos de análisis aplicable, ya que la pertinencia de este estudio es establecer el marco de la teoría de la CTS.

El IoT o "Internet de las Cosas", refleja la base (*sw + hw*) del proceso de convergencia tecno-social; donde la continuidad, disponibilidad y universalidad de la comunicación mediante el uso del "Internet" establecen la perene comunicación e interrelación (h + i + m); para (Italia, Authored, Minerva, Biru, & Rotondi, 2015), el IoT integra la tecnología con la sociedad; por esta razón el "Sistema Físico Cibernético", establece políticas de control y gobernanza a nivel de procesos, los cuales se encargan de establecer guías y mapas de ruta mediante "núcleos' en el arduo proceso de CTS. Dentro de la CTS, disponemos de la variable (m) "maquina", establece un enfoque de relación (*Hardware – Software*), el cual cubre objetos y estructuras en ambientes físicos-digitales estableciendo así la base de los "*Sistemas Físicos Cibernéticos*" como lo menciona (Ragunathan Rajkumar, 2010), esto nos permite entender de una manera más amplia los procesos de comunicación que conviven en una comunidad virtual, donde se establece la relación (h + i + m).

4.3 La cibercultura y las comunidades virtuales

El humanismo tradicional del siglo XX, inspirado por la cultura del libro, se distancia sistemáticamente de la nueva sociedad de la información digital como lo menciona (Chavarria, 2013), aunque el problema de fondo persista; los métodos cambian, en el periodo tipográfico las políticas de alfabetización se desarrollaron a nivel mundial, esto con el fin de establecer progreso social basado en un proceso de "Culturalización"; pero lamentablemente esto no erradico los problemas conductuales, sociales y económicos de nuestra sociedad (Coll, 2005); ya que el factor humano (h) es la clave en la transformación.

El siglo XXI empieza con marcados cambios sociales, políticos y económicos a nivel mundial, pero lo más relevante es la "Convergencia Tecno – Social" CTS; el cual establece una nueva forma de vida humana basada en la convivencia (h+i+m), dentro de un nuevo ecosistema físico-cibernético; estableciendo nuevas formas de comunicación humana que derivan en el Ciberantropo u hombre cibernético y en un nuevo tipo de unidades sociales llamadas "comunidades virtuales"

Aunque somos parte de un proceso evolutivo inmerso en el desarrollo de diferentes comunidades virtuales, las mismas establecen sus principios de convivencia en el uso de herramientas tecnológicas, esto sin duda, ha impactado en las unidades sociales como: la Familia, educación, iglesia, industria, gobierno y demás; desarrollándose vertiginosamente con características distintas las cuales se establecen bajo requerimientos tecnológicos específicos, siendo participes de una nueva generación de sociedades que habitan en el ecosistema físico-cibernético existente.

La convivencia y desarrollo adaptativo de los miembros de estas comunidades son estudiadas en la actualidad rumbo a la compleja y completa integración tecno-social, la cual ha llegado a impactar notablemente en la sociedad en algunas áreas como la seguridad pública, familia, iglesia, industria y educación; desarrollándose ciertos constructos como la influencia, el control, y la réplica de conductas dentro del ecosistema, de ahí el desarrollo de "*La perspectiva social del riesgo*" según (Tolosana, 2007), como uno de los factores de estudio dentro de la CTS.

Es importante considerar que las comunidades virtuales son evolutivas, disponen de constantes cambios y su interés es común; sus requerimientos tecnológicos permiten una interacción simétrica y ubicua con otros miembros de su colectivo; en este punto es importante agregar al vector masivo de impacto (i), el cual es un moderador de fenómenos vinculantes entre la tecnología y sociedad sugiriendo nuevas investigaciones en torno a los niveles de impacto de este vector con el uso de otros métodos investigativos como el cuantitativo o mixto.

4.4 La "información" (i) como vector masivo de impacto

Los colectivos sociales han desarrollado una nueva manera de gestionar la información, la cual es de impacto a todo nivel dentro de sus respectivas unidades sociales, estableciendo de este modo la importancia del "Pensamiento Colaborativo" también llamado en Ingles "*Collaborative Thinking*" (Pang & Lee, 2008), el cual conlleva un pensamiento amplio y global referente a la manera como el humano interactúa dentro del nuevo ecosistema físico-cibernético; siendo este un nuevo patrón en la manera de vivir del ciberantropo u hombre cibernético.

Este ecosistema trae en si un modelo colaborativo de participación, diversidad cultural e integración inherente a una forma colectiva de comunicación y pensamiento social colaborativo, ya que la individualización en la gestión de información es prácticamente nula, esto debido al entendible en el nuevo proceso de pensamiento humano, el cual rige, las redes sociales (Hillman, 2007).

Una de las principales características de la Convergencia "Tecnología Social" es la manera como se comparte la información, el uso del Web 2.0, el cual hoy en día ha estandarizado la manera de compartir información juegan un papel fundamental en el manejo de información (Italia et al., 2015). Ha todo esto es importante considerar que en este nuevo ecosistema físico-cibernético se gestiona la información en todos sus procesos, lo cual es vital para el respectivo uso y "diseminación" lo cual conlleva "compartir y guardar información".

En este proceso de organización de información la "Folcsonomia" es de suma importancia ya que este método nos permite clasificar, agrupar y organizar información para ser usada de una manera fácil a través de "Tags" (Hayman, Schemes, & Square, 2007), para poder entender su función se explicara la raíz etimológica de la palabra la cual está compuesta por Folk "gente, pueblo, nación familia, unidad" y Taxonomía "categorización", la misma que fue usada por primera vez por Thomas Vander Wal en el 2004 de acuerdo a (Knoll, 2006), este método de etiquetado social incrementa sustancialmente el proceso rápido de búsqueda, identificación y análisis de una palabra relevante asociada o asignada a un pieza de información (Hayman et al., 2007).

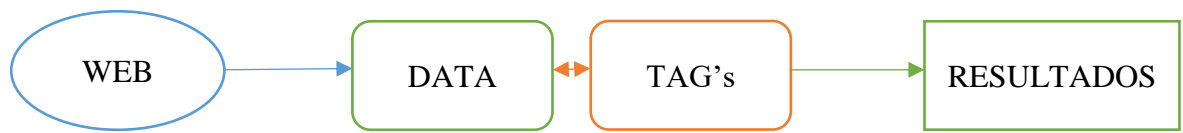

Figura 2 Estructura de los "Tags" usando la Folcsonomia (por: Rommel Salas)

La vinculación del uso de base de datos en conjunto con los servicios de consulta ofrecidos por las máquinas de búsqueda nos da como resultados un proceso ágil y sistemático al momento de buscar y compartir información, este proceso ya familiar y común en el periodo digital es la base del proceso informativo social el cual está sujeta en función de los resultados obtenidos; este proceso de gestión de información tiene un "Ciclo de Vida", el cual se describe a continuación:

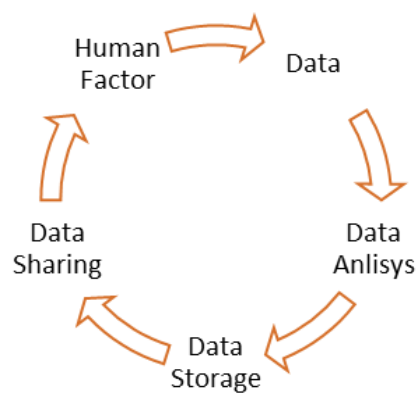

Figura 3 Ciclo de Vida de la Información en la Convergencia Tecno-Social (por: Rommel Salas)

Este clico de vida prioriza la gestión social de la información, la cual conlleva algunos procesos encaminados al uso de la información dentro de las unidades sociales, por esta razón desarrollaremos un mapa conceptual con el propósito de establecer los componentes de cada uno de los procesos mencionados en el grafico anterior (ver figura 3), El factor humano es la base de este proceso ya que encausa las "necesidades y responsabilidades" dentro de los componentes de las unidades sociales, luego observamos a la "Data" que dentro de su tipología esta la "información privada y pública"; es importante recalcar que dentro de la "privada" tenemos información personal que debe ser protegida y no expuesta dentro de la comunidad virtual; y la "publica" es información social, expuesta públicamente, la misma que debe estar filtrada por aspecto ciberéticos.

Esta Data es analizada sistemáticamente mediante las herramientas basadas en la cadena jerárquica de información y organizadas mediante métodos de etiquetación social; la misma que luego de ser analizada es almacenada mediante diferente tipos de procesos y unidades de almacenamiento fijos, móviles o virtuales; en este segmento la información culmina el ciclo en "InfoSharing", considerando las nuevas formas de interacción humana basada en (h+i+m), lo cual permite diseminar la información por diferentes medios como: redes sociales, herramientas colaborativas para luego volver al factor humano (h).

4.4.1 Cadena jerárquica de información DIC

Para entender la cadena DIC, debemos partir de la taxonomía del proceso de información (*Data, información, conocimiento*), la cual es la base de nuestra propuesta para entender el uso eficiente de la información "Sistema de Información Eficiente", desde su inicio, el conocimiento de esta jerarquía nos permitirá asociar los conceptos, logrando entender claramente el manejo correcto de la información usando la tecnología, este componente estratégico es el cual debe considerarse como una base epistemológica para nuevos estudios cuantitativos con el descubrimiento de constructos y paradigmas nuevos en la investigación científica.

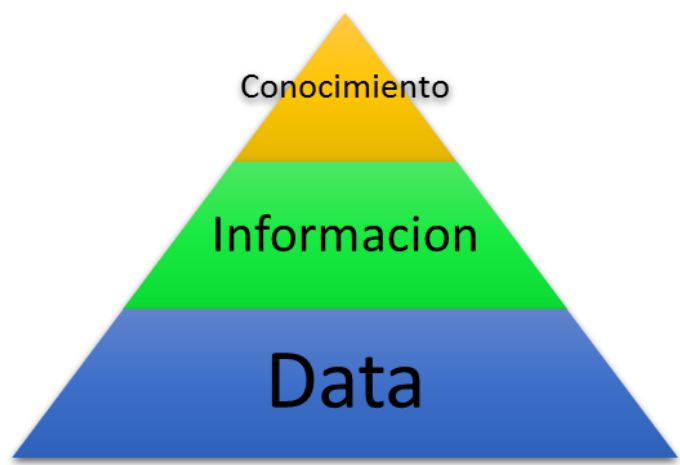

Figura 4 cadena jerárquica DIC (por: Rommel Salas)

4.4.2 Cadena jerárquica de información: los datos

A los Datos se le puede definir como la corriente de hechos en bruto que representan los acontecimientos que ocurren en los procesos de gestión de información social (Benavides & Tompkins, 2009), antes de que hayan sido procesadas y dispuestas en una forma **CUPEO** (*Completa, utilizable, precisa, exacta, oportuna*), para que sus respectivos componentes puedan entender y usar correctamente (Salas, 2015).

Como menciona (Hayajneh, 2013), los datos pueden recogerse de muchas fuentes, ya sean directa o indirectamente. Cuando hablamos de **recolección directa** nos referimos al proceso de recolección de data para un propósito específico sean estas necesidades o responsabilidades. Por ejemplo, "las horas en las que un reloj de ingreso de empleados donde la data es recogido a base de un proceso biométrico", y estos datos se utilizan para el pago de nómina. Por otro lado, la información puede derivarse de los datos que se recogieron originalmente para un propósito completamente diferente, esto es la **recolección indirecta**. Por ejemplo, una compañía de tarjetas de crédito recopilará datos sobre compras realizadas de manera virtual, con esta data la compañía, además, de procesar y organizar la información de estos clientes puede vender a una empresa de cruceros, para el desarrollo de una campana de mercadeo, con la recolección indirecta los datos son usados para un fin distinto de aquel para el que fue originalmente recogido.

4.4.3 Cadena jerárquica de información: la información

Son datos que han sido procesados con la finalidad de ser útiles. Esto es de vital importancia en la gestión diaria en los procesos de comunicación dentro de las unidades sociales, para (Christopher Connor, 2013) la información está destinada a cambiar la forma como el usuario que es el receptor percibe algo (para tener un impacto en su juicio o comportamiento, el cual como ya lo hemos mencionado anteriormente deriva en la "Perspectiva Social del Riesgo"), lo cual resultará en conocimiento, la información debe contener los siguientes atributos (Salas, 2015):

- **Completa:** Es la que proporciona todos los datos necesarios.
- **Utilizable:** Es de fácil comprensión para los formatos y pantallas.
- **Precisa:** Tiene el nivel de detalle requerido.
- **Exacta:** Es la que transmite la verdadera situación.
- **Oportuna:** Está disponible en el momento de tomar las decisiones.

Las principales falencias en los procesos de control, seguridad y calidad en los resultados de la funcionalidad de los sistemas de información dentro de las unidades sociales, conlleva al aparecimiento de diferentes fenómenos sociales, los cuales son discriminados por factores externos al "*procesamiento de la data*" por parte de diferentes elementos como el humano, el cual altera la cadena de procesamiento de información a nivel de afectar a las unidades sociales y por ende a sus nuevos componentes llamados comunidades virtuales.

4.4.4 Cadena jerárquica de información: el conocimiento

De acuerdo con (CEN, 2004), es la combinación de los datos y la información, a la que se añade la opinión de expertos, habilidades y experiencia, para dar lugar a un activo valioso dentro de los procesos de gestión de información social. El procesamiento de información de un sistema es sistemático, esto implica una transformación del conocimiento basado en el procesamiento de la información con el fin de realizar y completar funciones dirigidas por el humano. Para (Kampfner, 2010) este conocimiento es transformado, de hecho, está incluida en la estructura y la dinámica del DIC y se utiliza o se manifiesta a través de su dinámica, es decir, "los procesos que realiza".

4.4.5 El procesamiento de data a información

Luego de la adquisición o recolección de la data en diferentes fuentes, esta es procesada hasta llegar hacer útil, dentro de del origen de la palabra y partiendo del significado original del verbo "informar", el cual denota "dar forma" a algo. Por lo tanto, para que los datos se conviertan en información como manifiesta (Hey, 2004), estos deben tener forma o estructura para luego ser usadas por el receptor por medio de sus habilidades y experiencias.

Mediante este procesamiento la sustancia de datos en una forma útil se convierte en información, tal como materiales procesados pueden convertirse en un producto útil, estableciendo nuevos y útiles resultados (Salas, 2015). Con el uso de herramientas tecnológicas se reciben, procesan y comparten información (i) adecuadamente (Best, Cumming, & Se, 2007), más el uso correcto del factor humano (h) basado en sus necesidades y responsabilidades, es la que nos permitirá disponer de información (i) que cumpla con la forma **CUPEO** (*Completa, utilizable, precisa, exacta, oportuna*).

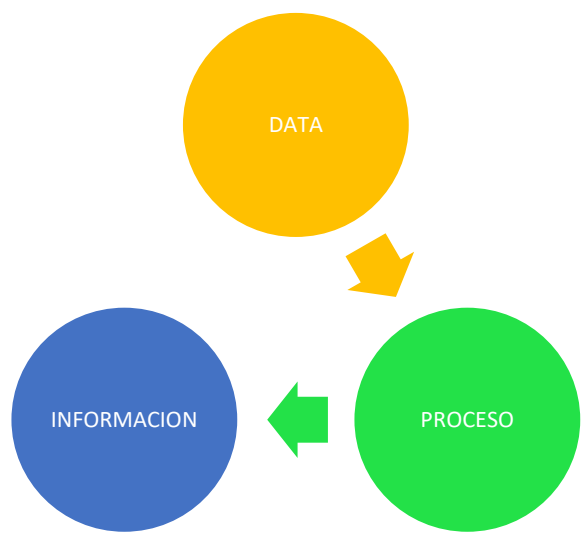

Figura 5 Transformación de la data (por: Rommel Salas)

4.4.6 El procesamiento de información al conocimiento

Luego de disponer información útil dentro de las unidades sociales, el correcto uso y gestión del factor humano, da como resultado que la información "un medio o material" sea necesario para obtener y construir el conocimiento (Hayajneh, 2013).

Para Nonaka (1996), dentro de su teoría establece que el conocimiento explícito puede ser almacenado y compartido mientras que el conocimiento tácito debe ser internalizado para informarnos sobre el "*know-how*". De ahí la importancia que tiene la transformación de data en información y los atributos luego de ser procesada (CUPEO).

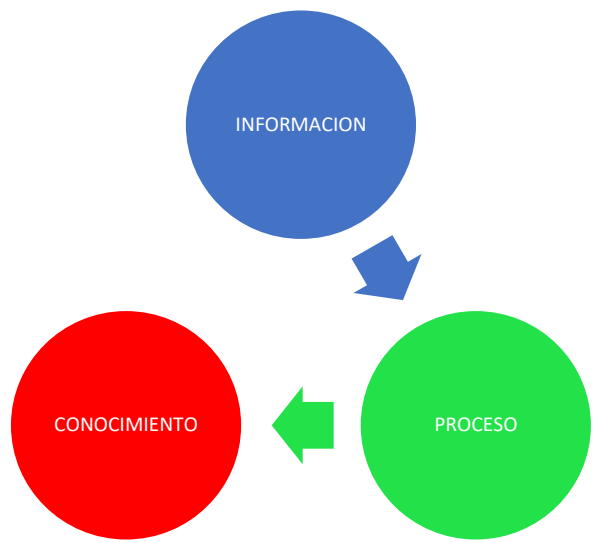

Figure 1 Transformación al conocimiento (por: Rommel Salas)

5. Cambios conductuales y cognitivos del humano dentro de la CTS

5.1 Impacto individual

De acuerdo a (Blute, 2005), la "Memética", es el estudio científico de los "memes", el cual fue impulsado por Richard Dawkins, la misma que está compuesto en dos partes "m" *memoria o imitación*, y "eme" *genes*; lo cual como menciona este autor es la unidad básica de información hereditaria o replicación involucrada en actividades culturales.

Cuando partimos de la premisa que el aprendizaje social por observación implica aspectos sensoriales (Myers, 2005), los cuales a menudo son instantáneos elaborándose sobre el pensamiento deliberado, el cual lleva información del ojo u oído al cerebro, específicamente al "Tálamo"; y hacia la "Amígdala" la cual es centro del control emocional del ser humano, y esto antes que el pensamiento intervenga, nos permite establecer un factor isomorfo del proceso de impacto conductual, el cual afecta a las unidades sociales dentro de la convivencia tecno-social, por lo tanto, podemos entender así la importancia de los niveles de estructura de los [m-eme], los cuales están compuestos en: Memotipo, Mediotipo y Sociotipo, como lo menciona (Rose, 1998).

- **Memotipo:** describe la información grabada en la memoria.
- **Mediotipo:** describe la información en la forma como se expresa.
- **Sociotipo:** describe donde la información es grabada en la memoria de los individuos.

El Ciclo de Vida del "Meme" acorde a dicho autor es:

- **Asimilación:** envuelve la noticia, entendimiento y aceptación, donde el humano es impactado por observación de otros individuos, "Leyendo, Escuchando, para generar así una nueva idea"; la asimilación corresponde al "Nivel Mediotipo".
- **Retención:** luego de la asimilación, esta se retiene en la memoria, este es un Nivel de Memotipo, ya que está envuelto procesos cognitivos, (Aquí se debe evaluar de qué manera contrastan las reglas y normas culturales y la influencia basada en aprendizaje cognitivo). Por esta razón la Información que evoca la emoción deja una fuerte huella en la memoria como lo menciona (Phelps, 2004).
- **Expresión:** el proceso de propagación y replica desde un humano a otro, se basa en la transformación y trasmisión de la información; estas imitaciones pueden basarse en el hablar, la escritura, las acciones (Pempek, Yermolayeva, & Calvert, 2009).
- **Transmisión:** Está basada en un nivel de Mediotipo, ya que es la forma como se expresa la información, si esta es corrupta o distorsionada, durante la transmisión el resultado tendrá muchas variaciones de información (Pang & Lee, 2008).

Es importante considerar que la imitación e influencia dentro de las redes sociales por medio de la información generada es un factor de impacto (Feldman, 2010), al cual le denominamos "vector masivo de impacto" por esta razón, aunque dos gemelos idénticos dispongan de la misma genética, van a manifestar distintos patrones de desarrollo las cuales son atribuidas por las variaciones en ambientes de crianza del individuo. Se debe considerar los enunciados científicos relevantes a los estudios basados en personas influenciadas por ambientes similares, pero con genética distinta, para entender de esta manera que tanto factores extrincicos, como intrincicos.

La red social es definido por (Lazega, 1998), como "*un conjunto de relaciones especificas – apoyo, consejo, control o influencia entre número limitado de actores*"; esta interrelación colaborativa basada en comunidades virtuales dedicadas a compartir información, con el uso de tecnología especifica basada en hardware, software y transportada mediante internet son las que comúnmente se les conoce como redes sociales que actualmente están disponibles dentro de las comunidades virtuales emergentes en la sociedad actual.

La información que es transmitida mediante el "Mediotipo" se aloja en la memoria la cual de acuerdo a (Feldman, 2010) es la habilidad para codificar, almacenar y recuperar información. En este punto es importante analizar qué información se transmite, como se filtra, y lo más importante cuales pueden ser las consecuencias en la transmisión de información que no solamente sea errada sino altere o motive a la persona a un cambio que afecte su entorno.

El impacto en las unidades sociales puede ser alarmantes cuando la información es corrupta o distorsionada dentro del proceso de comunicación, algo muy común en las redes sociales, y en las máquinas de búsqueda como google, la influencia y control es eminente ya que sus cualidades ejecutorias son controladas (Daniel papalia, 2010) dentro de la función ejecutiva, la cual va acompañada al desarrollo del cerebro, específicamente a la "*Corteza Prefrontal*"; en esta región se desarrolla la planeación, el juicio y la toma de decisiones.

5.2 Impacto colectivo

El uso de la tecnología desde su inicio impacto profundamente a todas las esferas de poder como menciona (Ortiz, 2008), "la historia de las tecnologías de la información y la comunicación, está estrechamente ligada a procesos sociales, tanto de poder como de contrapoder, de dominación y de resistencia", dando un énfasis importante a los procesos de comunicación masiva como en el caso de los partidos políticos, los cuales usan la tecnología para un proceso impositivo dogmático, favoreciendo así a las comunicaciones en masa como se hace la referencia en el mencionado trabajo.

Este "enfoque entre el poder y contrapoder, en la política formal, en la política insurgente y en las masas sociales" de (Ortiz, 2008), establece algunas consideraciones y experiencias político sociales de Colombia, un país que por más de medio siglo ha vivido sumido en constantes conflictos políticos y sociales, esta experiencia explica características de un "entorno altamente complejo, socializador y socializado de convergencia de viejos y nuevos medios y tecnologías, donde se configura una novedosa cibercultura", la cual tiene como énfasis una población impactada masivamente por medios tecnológicos como un detonador intrínseco en los procesos de impacto ideológico político masivo.

Dentro del análisis de la convivencia tecno-social, sus componentes están completamente integrados en base a los objetivos del proceso de manejo de información, la cual como menciona (Kendall, 1998), cambia y transforma las entradas y salidas de información haciendo un análisis minucioso de cómo este evalúa los cambios sustanciales en la información procesada, mediante la verificación y actualización de información.

Este proceso en el manejo de información se manifiesta en el uso de información dentro de las unidades sociales con interrelación e interdependencia en aspectos humanos del diario vivir, donde los componentes de estas unidades sociales participan, interpretan, replican y transcienden dentro de sus comunidades virtuales, permitiendo de esta manera el desarrollo de nuevos procesos de expansión de comunicación ubicua.

6. El Marco Conceptual de la Investigación (CTS Framework)

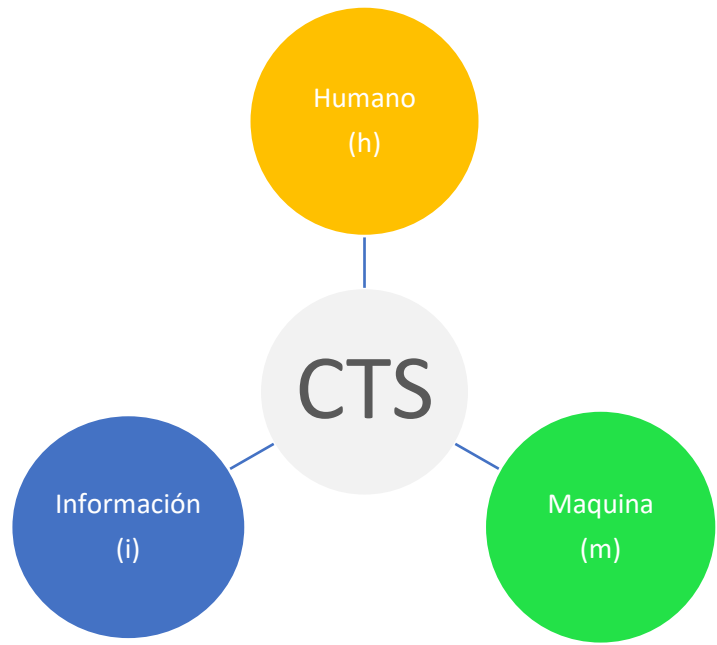

7. Metodología

Los nuevos procesos epistemológicos que emergen en el Siglo XX descritos por (Martínez, 2006) en dos componentes sustanciales: a) el investigador y su sensibilidad en cuanto a los métodos, técnicas, estrategias y procedimientos; b) el investigador y su rigor, sistematicidad y criticidad; y la influencia de la escuela de chicago, como lo menciona (Piovani, 2011), favorecen al desarrollo de métodos basados en documentación personal, trabajo de campo, análisis de fuentes documentales y mapeo social; uno de los legados de esta escuela es la "teoría fundada" la cual de acuerdo a (Universidad Alberto Hurtado, 1995) descubre conceptos, hipótesis y relaciones a partir de los datos para la comprensión y explicación de fenómenos sociales.

De ahí que las nuevas investigaciones de la cibernética social basadas en fenómenos observados por los investigadores sociales carecen de cierta rigidez metodológica; por esta razón los métodos investigativos usados en el área de la cibernética social dentro de la academia, suelen establecerse de una manera ambivalentemente con una fuerte base en la intencionalidad, de este modo vamos a establecer el método científico a usar, pero sin antes explicar estos dos conceptos dicotómicos.

De acuerdo a (Quintana, 2006) el método cuantitativo se centra en la explicación y predicción; y el método cualitativo en la comprensión de una realidad considerada desde los aspectos particulares; continuando con esta línea de pensamiento social cualitativo; para (Ruiz Olabuénaga, 1996) este método estudia la realidad en su contexto natural, tal y como sucede, de ahí la importancia del establecimiento del mismo para poder entender y describir los fenómenos sociales de la convergencia tecno-social desde su interior; permitiéndonos de este manera establecer una nueva dirección, con particulares características para el desarrollo de la investigación en la Antropología de la Informática Social.

7.1 Rigor metodológico

El estudio de los nuevos fenómenos cibernéticos dentro del área social, requieren métodos rigurosos y sistematizados en el desarrollo de conocimiento científico social, el enfoque conductual y el uso de tecnología nos lleva a innovar procesos que se ajusten a la realidad en la que vivimos (Ecosistema h+i+m); y no simplemente a la aplicabilidad de la influencia de las escuelas cuantitativas de otras disciplinas, las cuales ejercen cierto predominio en el área social sin ser parte de la esencia en los estudios de la Antropología de la Informática Social.

Este problema tiene un matiz epistemológico, con vagas definiciones ilusas influenciadas por la crisis cualitivista del siglo pasado; que en muchos de los casos crearon la estigmatización de un método cualitativo sin rigor y sistematización (Glaser & Strauss, 1967); esto nos lleva a un proceso de estancamiento en el desarrollo de teorías en las áreas del cibernética social; ya que como menciona (Martínez, 2006) nuestra mente no sigue únicamente un enfoque causal, lineal, unidireccional; sino también modular, estructural, dialectico, gestáltico, interdisciplinario, donde todo es afectado por todo, y esto es parte del periodo digital donde nos desarrollamos, no de una manera autónoma; sino, en armonía con otros componentes de este nuevo ecosistema; esto es la red de relaciones con los demás, la base donde se envuelven los nuevos fenómenos cibernéticos sociales.

Dentro del campo social el método cualitativo sigue siendo la única forma de obtener datos sobre muchas áreas donde convergen los fenómenos sociales (Glaser & Strauss, 1967), ya que estas unidades sociales interactúan dinámicamente; la dicotomía científica ha generado una brecha existente basada en la limitación de la recolección de datos cuantitativos, impulsando a establecer parsimoniosamente nuevas iniciativas en la investigación científica cualitativa que envuelvan al investigador como parte principal dentro de los nuevos fenómenos a investigar en el ecosistema físico-cibernético.

7.2 Base epistemológica tradicional cualitativa y la cibernética social

7.2.1 Los métodos y las técnicas en la investigación social

Dentro del rigor metodológico; es importante describir el concepto de método, (Poncela, 1990) manifiesta que son los procedimientos y pasos concretos seleccionados según el objeto y fin de la investigación; de igual manera para (López-Herrera & Salas-Harms, 2009) el método, en la investigación social es el procedimiento lógico que el investigador debe seguir para acercarse a la verdad y verificarla.

En atención a lo expuesto anteriormente, las técnicas en la investigación social están definidas por (Poncela, 1990), como un conjunto de aspectos operativos basados en la organización de datos, la formación de conceptos e hipótesis, recopilación e análisis de información y producción de resultados; de igual modo (Palomino, Alonso, Morales, & Prieto, 2009) establece el concepto de técnicas como procedimientos operativos rigurosos, bien definibles, transmisibles para repetirse sucesivamente, el cual depende del objetivo que el investigador busca y su método de trabajo.

7.2.2 El método cualitativo en la investigación social cibernética

Para (Universidad Alberto Hurtado, 1995), la investigación estratégica cualitivista, en diferencia a la cuántica cuantitavista, tiene un proceso secuencial y distributivo basado en la adaptación de las características particulares de aquello que se va a estudiar. Para (Quintana, 2006) este método investigativo se centra en la comprensión de una realidad vista desde los aspectos particulares como fruto de un proceso social; el cual, dentro de la CTS, se puede establecer que no existe más una separación entre el humano (h), la información (i) y la maquina (m), ya que estos conviven dentro del nuevo ecosistema físico-cibernético y por ende es una realidad vista desde adentro de las nuevas unidades sociales llamadas comunidades virtuales.

Para (Ruiz Olabuénaga, 1996) el método cualitativo estudia la realidad en un contexto natural, tal y como sucede, este concepto nos lleva a entender que los estudios basados en la cibernética social, deben realizarse dentro de una comunidad virtual, lo cual implica que el investigador este envuelto en los fenómenos de estudio. Este es un enfoque que establece criterios significativos para las personas implicadas en el ámbito de estudio; ya que para (López-Herrera & Salas-Harms, 2009) el método cualitativo se caracteriza por evitar la cuantificación para basarse en descripciones narrativas, estudiando los contextos estructurales y situacionales, donde esas vivencias diversas, marcan los nuevos hitos del ciberantropo.

8. Implicaciones cibernética social

La Teoría de la CTS, nos permite identificar, describir y establecer cuáles fueron los efectos del periodo digital en las unidades sociales, así como el establecimiento de un nuevo ecosistema de convivencia humano-maquina; conjuntamente con los futuros estudios de los niveles de impacto basados en; influencia, control y replica; mediante el vector masivo (i) "información" de impacto, además la falta de criterio ambivalente de los órganos de control social hacia el establecimiento de normas Ciberéticas que establezcan el fundamento de la forma de vida del nuevo Ciberantropo. El camino es largo por recorrer, tenemos a la CTS que establece la ontología de la cibernética moderna, y sus respectivos enunciados epistemológicos, su aplicación es interdisciplinaria y multidisciplinaria.

8.1 Efectos e impacto del periodo digital en las unidades sociales

Dentro del estudio de la convergencia tecno-social encontramos un enfoque de la comunicación social digital, que agilizo el desarrollo de un proceso de convergencia (h + i + m);  con el establecimiento de una manera nueva de compartir información dinámicamente, como es el caso de la Web 2.0, el cual dispone de infraestructura dinámica colaborativa en diferencia con sus antecesores, los cuales disponían de un modo estático informativo, como lo manifiesta (Frydenberg, 2011), paulatinamente el Web 2.0 ha transformado las redes sociales, y el trabajo en grupo, esto ha permitido establecer una variedad de servicios que disponen en la actualidad las nuevas páginas web y por ende las comunidades virtuales las cuales se nutren de los procesos colaborativos de este enfoque de comunicación que es la base sistémica de la CTS.

Por este motivo como lo menciona (Levy, 2006), el desarrollo de la administración de información ha llegado a niveles donde las nuevas técnicas de "*Análisis de Información*" como la: Minería de Datos, Data Warehouse, Análisis de Información en Redes Sociales, y su aplicación a todas las áreas de la sociedad dentro de las herramientas colaborativas, establecen una nueva manera de generar conocimiento.

El desarrollo de aplicaciones basadas en soluciones corporativas, educativas, conductuales y más; dan un impulso hegemónico al vínculo "Tecno-Social", impactando a todas las unidades sociales, dentro de las cuales empiezan a desarrollarse las áreas "*vulnerabilidad social*"; las cuales establecen una brecha dentro de la trilogía (h + i +m); ya que los niveles de impacto afectan las variables (h) humano y (m) máquina, de una manera aleatoria.

Dentro de este estudio podemos entender que muchos de los beneficios del impacto tecnológico y la gestión de información en la sociedad, y su convergencia han llegado con un sinnúmero de fenómenos que se han transformado en "*Riesgo*" para los gobiernos y sus ciudadanos, para futuras investigaciones quedarían pendientes preguntas como: "El análisis de los impactos de la convergencia  "Tecno-Social" en sociedad", Modelos para medir el impacto Humano – Maquina; lo cual nos permitirá entender el comportamiento de ciertos grupos sociales luego de la llegada de la convergencia y como estos mutaron a comunidades virtuales emergentes. Lo que si podemos afirmar; luego de completado nuestro estudio es que la sociedad cambio, se estableció nuevas formas de comunicarse, lo que ha impactado en todas las áreas de la sociedad.

El nivel de convergencia es completo, a todos sus niveles dentro de las unidades sociales; vemos como ejemplo los casos de la inteligencia de negocios o bioinformática, la cual como menciona

(Thacker, 2007) disponen de un medio tecnológico con particular énfasis en la simulación, modelo y análisis del origen de data biológica, así también el estudio del ADN Informático (Thacker, 2007) el cual de acuerdo al mencionado autor contiene tres elementos cruciales disponibles en dispositivos electrónicos el procesamiento, la gestión y el almacenamiento.

Como vemos la convergencia nos ha llevado a un proceso de indexación tecno-social completa (Guerrera, 2011), donde este proceso va un paso más allá de las personalizaciones individuales ya que incorpora procesos de "colaboración e *InfoSharing*" y la respectiva información pública y privada de las personas con las que comúnmente estamos conectados.

8. Conclusión: Lo relativo a una nueva convivencia h+i+m

Este impacto ha establecido el desarrollo de investigaciones basadas en actuales necesidades de la sociedad, dentro del cual la mayoría de las unidades sociales carecen de protocolos que ejecuten normativas y estándares que faciliten el uso de la tecnología y el manejo de la información para una convivencia sana, este es un punto determinante en el comportamiento humano basado en el uso de la tecnología ya que la falta de educación sistemática y el descuido en la implantación de directrices "Ciberéticas" para el uso correcto de la tecnología luego de la convergencia empieza a dar sus primeros frutos, que es la brecha generacional tecnológica, la cual separa generaciones dentro de las unidades sociales.

El Fenómeno BGT (Brecha generacional tecnológica), establece modos de conducta generacional (Richard Gelles, 2000), este comportamiento social cotidiano basado en el interaccionismo simbólico "supone que el comportamiento humano no es determinado por los hechos objetivos de una situación, sino por el significado que las personas atribuyen a dicha situación." La aplicación y uso de la tecnología es no solamente complementaria sino "Integradora"; ya no podemos hablar de dos áreas separadas, con la convergencia "Tecno-Social" se unifico ahora interactúan en conjunto, a lo largo de los años el manejo de la información ha impactado la historia de la civilización humana, esto fue el caso de la escritura, la imprenta, y la infotecnología.

Hoy en día el uso es común, indispensable e irreversible; ¿qué sucedería si una ciudad no tiene computadores, teléfonos celulares, internet, correos electrónicos, mensajes de texto?, y si vamos más profundamente el uso de los dispositivos de procesamiento de información en la industria de servicios como la: Luz eléctrica, agua potable, salud, transporte y educación; sería catastrófico, esta es la realidad de nuestra nueva sociedad basado en el ecosistema físico-cibernético.